\documentclass[twocolumn,pre,twoside,floatfix]{revtex4}

\usepackage{amsmath}
\usepackage{amssymb}
\usepackage{graphicx}
\usepackage{dcolumn}
\usepackage{bm}
\usepackage{color}

\def\dps{\displaystyle}

\def\Eq#1{(\ref{eq:#1})}
\def\d{\mathrm{d}}

\def\epsilon{\varepsilon}
\def\theta{\vartheta}
\def\rho{\varrho}

\def\vec#1{\mathbf{#1}}

\def\div{\mathop\mathrm{div}}

\begin{document}


\title{Non-equilibrium interfaces in colloidal fluids}

\author{Markus Bier}
\email{bier@is.mpg.de}
\author{Daniel Arnold}
\affiliation
{
   Max-Planck-Institut f\"ur Intelligente Systeme, 
   Heisenbergstr.\ 3,
   70569 Stuttgart,
   Germany, 
   and
   Institut f\"ur Theoretische Physik IV,
   Universit\"at Stuttgart,
   Pfaffenwaldring 57,
   70569 Stuttgart,
   Germany
}

\date{11 September 2013}

\begin{abstract}
The time-dependent structure, interfacial tension, and evaporation
of an oversaturated colloid-rich (liquid) phase in contact with an 
undersaturated colloid-poor (vapor) phase of a colloidal dispersion is 
investigated theoretically during the early-stage relaxation, where
the interface is relaxing towards a local equilibrium state while 
the bulk phases are still out of equilibrium.
Since systems of this type exhibit a clear separation of colloidal and solvent
relaxation time scales with typical times of interfacial tension measurements
in between, they can be expected to be suitable for analogous experimental
studies, too.
The major finding is that, irrespective of how much the bulk phases differ
from two-phase coexistence, the interfacial structure and the interfacial
tension approach those at two-phase coexistence during the early-stage
relaxation process.
This is a surprising observation since it implies that the relaxation towards
global equilibrium of the interface is not following but preceding that of
the bulk phases.
Scaling forms for the local chemical potential, the flux, and the dissipation
rate exhibit qualitatively different leading order contributions depending on
whether an equilibrium or a non-equilibrium system is considered.
The degree of non-equilibrium between the bulk phases is found to not 
influence the qualitative relaxation behavior (i.e., the values of power-law 
exponents), but to determine the quantitative deviation of the observed 
quantities from their values at two-phase coexistence.
Whereas the underlying dynamics differs between colloidal and molecular fluids,
the behavior of quantities such as the interfacial tension approaching the 
equilibrium values during the early-stage relaxation process, during which 
non-equilibrium conditions of the bulk phases are not changed, can be expected
to occur for both types of systems.
\end{abstract}

\maketitle


\section{Introduction}

Interfaces between two fluid phases are commonly characterized by the 
interfacial tension, because this quantity is sensitive to the interfacial
structure \cite{Dietrich1988,Rowlinson2002} and, at the same time, it is easy 
to measure by means of numerous experimental methods developed during the last 
two centuries \cite{Drelich2006}.
Whereas the interfacial tension is thermodynamically defined only for 
two-phase coexistence \cite{Rowlinson2002}, it is frequently discussed also
for non-equilibrium conditions, e.g., in the context of the adsorption dynamics
of surfactants \cite{Millner1994, Eastoe2000}.
Since interfaces typically extend over a spatial range of the order of the
bulk correlation length, whereas the bulk phases are typically of macroscopic
extension, one can intuitively expect the existence of a time scale 
$\tau^\times$ such that at times $t\ll\tau^\times$ the interface relaxes towards
\emph{some} local equilibrium state the properties of which are governed by
the adjacent non-equilibrium bulk phases and at times $t\gg\tau^\times$ the
bulk phases relax towards the global equilibrium state.
In the following the processes taking place at times $t<\tau^\times$ or
$t>\tau^\times$ are referred to as the ``early-stage'' or the ``late-stage'' 
relaxation, respectively.
The late-stage relaxation processes have been extensively studied in the past,
probably because here one can expect the notion of a non-equilibrium 
interfacial tension being applicable \cite{Defay1977, Kayser1986, Lipowsky1986,
Peach1996, Lukyanov2013}.
However, neither the early-stage relaxation process of the interface nor the 
nature of the local equilibrium state of the interface in between two slowly 
relaxing non-equilibrium bulk phases are understood so far.
Obviously, at late times, when the ultimate global equilibrium state is 
reached, the interfacial structure is that in between two bulk phases at 
coexistence. 
However, for non-equilibrium bulk phases it cannot even be expected a priori
that the interface which has been equilibrated locally during the early-stage
relaxation bears a resemblance to any structure found at global equilibrium.
Moreover, one has to reckon with a non-trivial dependence of the local 
equilibrium interfacial structure on the properties of the adjacent 
non-equilibrium bulk phases.
In the present work we analyze these open questions on the early-stage 
relaxation process.

From the experimental perspective it is difficult to study the time-dependence
of the interfacial tension during the early-stage relaxation process of an
interface in a molecular fluid, because it relaxes on time scales much
shorter than the time of measurements of the interfacial tension.
One possible solution is to consider molecular fluids close to a critical point
\cite{Peach1996} or a wetting transition \cite{Kayser1986, Lipowsky1986}, 
which, however, is technically demanding due to the requirement of a stable 
temperature control.
Here it is proposed to alternatively study the time-dependence of the 
interfacial tension by means of colloidal fluids which can phase-separate 
into a colloid-rich (liquid) phase and a colloid-poor (vapor) phase 
\cite{Aarts2004b}.
The interfacial tension in colloidal fluids is typically some orders of 
magnitude smaller than in molecular fluids and the relaxation times can
be of the order of hours \cite{Aarts2004b}.
The clear separation of time scales in long relaxation times of the colloidal
structure, intermediate times to measure the interfacial tension, and short 
relaxation times of the solvent leads to a proper account of a time-dependent
interfacial tension for colloidal fluids.
For such systems the time-dependent sedimentation and interface formation in 
the gravitational field \cite{Aarts2004b} as well as the interface fluctuation
dynamics \cite{Aarts2004a} have been recorded using direct microscopy imaging.
However, these studies were devoted to the interface formation driven by an
external field.

In the present theoretical investigation we study the early-stage formation of
an interface between an oversaturated colloidal liquid in contact with an 
undersaturated colloidal vapor in the absence of external fields.
The analogous situation in a molecular fluid would lead to the evaporation of
the liquid; hence, the same term is used for colloidal fluids here. 
It should be stressed, that, although the colloids are dispersed in a molecular
solvent, the latter is assumed to be uniformly distributed throughout the 
system, whereas the interface is formed by the contact of the colloid-rich and
the colloid-poor phases.
Here only times longer than the relaxation time of the molecular solvent are
considered, for which the colloidal particles are in local equilibrium with the
solvent and exhibit diffusive Brownian dynamics.
An appropriate theoretical approach for this situation is dynamic density
functional theory (DDFT) \cite{Kawasaki1994, Dean1996, 
MariniBettoloMarconi1999, MariniBettoloMarconi2000}, which describes the 
conserved dynamics of time-dependent number density profiles of colloidal 
particles \cite{Evans1979, Dieterich1990}. 
Processes occurring on time scales shorter than the relaxation time of the 
molecular solvent, e.g., in the ballistic regime, are not considered in the
following.
The model of colloidal fluids used here, the formalism of DDFT, and the 
relevant observables, in particular the interfacial tension, are introduced
in Sec.~\ref{sec:formalism}.
The results on the interfacial structure dynamics, the time-dependence of the
interfacial tension, the energy dissipation during the interface formation as 
well as on the evaporation rates are discussed in Sec.~\ref{sec:results}.
Conclusions and a summary are given in Sec.~\ref{sec:conclusions}.


\section{\label{sec:formalism}Formalism}

\subsection{\label{subsec:model}Model}

Consider a three-dimensional dispersion of colloidal hard spheres of diameter
$d$ which in addition interact via the square-well potential
\begin{align}
   U(r) = 
   \begin{cases}
   -\epsilon & \text{, $r \leq \lambda$} \\
   0         & \text{, $r >    \lambda$}
   \end{cases}
   \label{eq:U}
\end{align}
with the cohesive energy $\epsilon>0$ and the extension $\lambda$ of the
attractive well.

In the following the focus is on collective properties of the colloidal fluid
which can be expressed in terms of the local number density $\rho(\vec{r})$ at
position $\vec{r}$. 
However, instead of describing the fluid structure by the number density
$\rho(\vec{r})$ it is more convenient to use the dimensionless occupancy 
$\phi(\vec{r}):=\rho(\vec{r})/\rho_\text{max}\in[0,1]$, where $\rho_\text{max}$
is the maximal number density of the fluid phase at the given temperature.
 
\begin{figure}[!t]
   \includegraphics{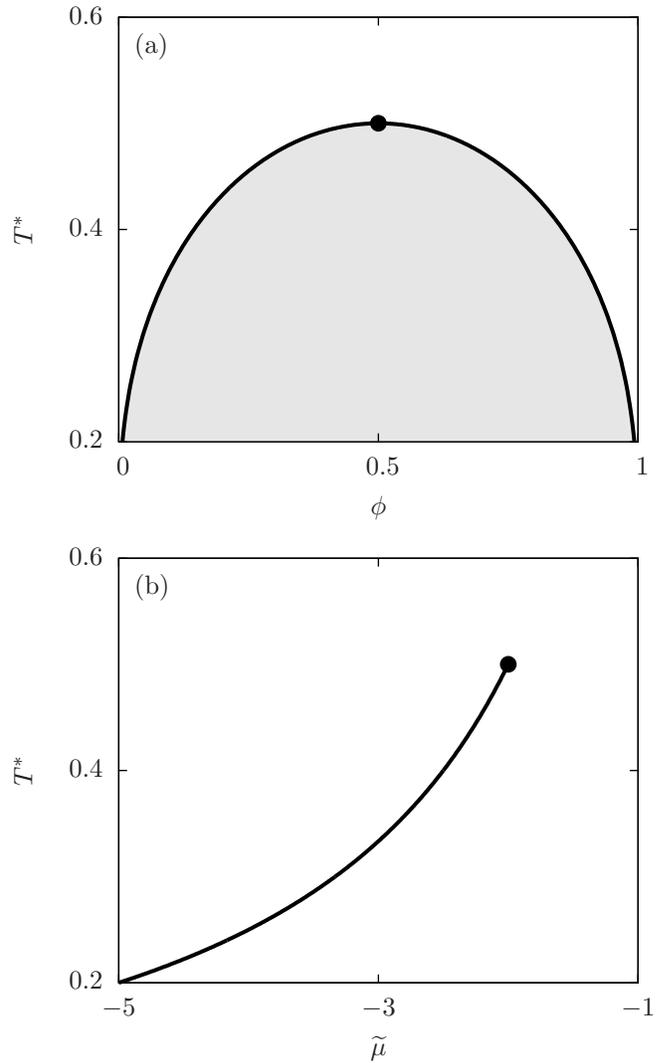}
   \caption{Phase diagram of the colloidal fluid described in 
            Subsec.~\ref{subsec:model} of hard spheres interacting by an
            additional square-well potential.
            The solid lines represent the liquid-vapor coexistence curves
            $T^*_\text{b}(\phi)$ (a, see Eq.~\Eq{binodal}) and 
            $\widetilde{\mu}_\text{co}(T^*)$ (b, see Eq.~\Eq{muco}).
            The two-phase coexistence region terminates in a critical point
            ($\bullet$, see Eq.~\Eq{criticalpoint}).}
   \label{fig:pd}
\end{figure}

Local equilibrium is assumed in the following, since only times longer that the
relaxation time of the molecular solvent are considered.
Consequently the temporal evolution of the system is described in terms of a
Helmholtz free energy density functional $F[\phi]$.
The present study does not aim for a quantitative description of a real 
colloidal fluid but for a qualitative and generic account of the early-stage
interface formation. 
Hence, in order to minimize the technical efforts, a free energy density 
functional 
\begin{align}
   F[\phi] = F_\text{HS}[\phi] + F^\text{ex}[\phi]
   \label{eq:df3d}
\end{align}
in the spirit of Ebner, Saam, and Stroud \cite{Ebner1976} is chosen, which 
comprises a \emph{local} free energy density functional $F_\text{HS}[\phi]$
describing a hard-sphere fluid and a \emph{quadratic} excess free energy 
functional $F^\text{ex}[\phi]$ to account for the square-well potential $U(r)$
(see Eq.~\Eq{U}).
Using a local functional $F_\text{HS}[\phi]$ to describe hard spheres is 
obviously not expected to be quantitatively precise, in particular in the
context of steep interfaces, and there are numerous highly sophisticated 
density functionals which reproduce the structure of hard-sphere fluids 
well \cite{Rosenfeld1989, Roth2002, HansenGoos2009}.
However, this is not expected to influence the general conclusions of this
investigation, because $F_\text{HS}[\phi]$ leads to qualitatively correct
profiles.

Here we chose the lattice-gas-like local hard-sphere functional
\begin{align}
   F_\text{HS}[\phi] 
   &= k_BT\rho_\text{max}\int\!\d^3r\;\big[\phi(\vec{r})\ln(\phi(\vec{r})) 
   \label{eq:Fhs} \\
   &\hphantom{= k_BT\rho_\text{max}\int\!\d^3r\;\big[}\
     + (1-\phi(\vec{r}))\ln(1-\phi(\vec{r}))\big] 
   \notag
\end{align}
and the excess functional within random phase approximation (RPA)
\begin{align}
   F^\text{ex}[\phi] =  
   \frac{\rho_\text{max}^2}{2}\int\!\d^3r\!\int\!\d^3r'\;
   U(|\vec{r}-\vec{r'}|)\phi(\vec{r})\phi(\vec{r'}).
   \label{eq:Frpa}
\end{align} 
The former is used, because it is the simplest form which leads to fluid 
densities $\rho(\vec{r})$ strictly in the interval $[0,\rho_\text{max}]$.
Moreover, the theoretical phase diagram resulting from 
Eqs.~\Eq{df3d}--\Eq{Frpa} (see Fig.~\ref{fig:pd} and
Subsec.~\ref{subsec:phases}) can be calculated analytically, and it agrees
semi-quantitatively with the fluid parts of the phase diagrams of real
colloidal dispersions \cite{Aarts2005}.
The perfect symmetry of the lattice-gas model with respect to an exchange of
particle and vacancy is irrelevant in the present study.


\subsection{\label{subsec:phases}Phase behavior}

Assuming a uniform bulk state the phase behavior of the colloidal fluid
introduced in the previous Subsec.~\ref{subsec:model} can be inferred from the
Helmholtz free energy density
\begin{align}
   f = 
   \frac{F}{V} = 
   k_BT\rho_\text{max}\left(\phi\ln(\phi) + (1-\phi)\ln(1-\phi) 
                            - \frac{\phi^2}{T^*}\right)      
   \label{eq:Fbulk}
\end{align} 
with the effective temperature
\begin{align}
   T^* = 
   \left(-\frac{\rho_\text{max}}{2k_BT}\int\!\d^3r\;U(|\vec{r}|)\right)^{-1} =
   \frac{3k_BT}{2\pi\epsilon\rho_\text{max}\lambda^3}.
   \label{eq:tstar}
\end{align}
The equation of state reads
\begin{align}
   \frac{p}{k_BT\rho_\text{max}} = -\ln(1-\phi) - \frac{\phi^2}{T^*}
   \label{eq:eos}
\end{align}
and the liquid phase coexists with the vapor phase at chemical potential 
\begin{align}
   \widetilde{\mu}_\text{co}(T^*) :=
   \frac{\mu_\text{co}(T^*)}{k_BT} = 
   \frac{\partial}{\partial \phi}\frac{f}{k_BT\rho_\text{max}}\Big|_\text{co} = 
   -\frac{1}{T^*}
   \label{eq:muco}
\end{align}
below the critical point ($T^* < T^*_\text{c}$), which is located at
\begin{align}
   T^*_\text{c}=\frac{1}{2}, \quad 
   \phi_\text{c}=\frac{1}{2}, \quad
   \widetilde{\mu}_\text{c} = \widetilde{\mu}_\text{co}(T^*_\text{c}) = -2.
   \label{eq:criticalpoint}
\end{align}
Finally the liquid-vapor binodal curve is given by
\begin{align}
   T^*_\text{b}(\phi) = 
   \frac{2\phi-1}{\dps\ln\left(\frac{\phi}{1-\phi}\right)}.
   \label{eq:binodal}
\end{align}
The phase diagram is displayed in Fig.~\ref{fig:pd}.
Due to the choice of a simple lattice-gas-like description the phase diagram
in Fig.~\ref{fig:pd}(a) is symmetric with respect to the critical occupancy
$\phi=\phi_\text{c}$.
However, this symmetry is irrelevant to all conclusions to be drawn later.


\subsection{Dynamic density functional theory}

If the system is prepared in an arbitrary initial state $\phi(\vec{r},t=0)$, its
state $\phi(\vec{r},t>0)$ evolves with time $t$ such that the Helmholtz free
energy $F[\phi(t)]$ reaches a minimum at $t\to\infty$.
In the present work the colloidal processes to be described in terms
of $\phi(\vec{r},t)$ are much slower than the molecular degrees of freedom.
Hence one can assume local thermodynamic equilibrium and define the local
chemical potential \cite{Dieterich1990}
\begin{align}
   \mu(\vec{r},[\phi(t)]) = 
   \frac{1}{\rho_\text{max}}\frac{\delta F}{\delta \phi(\vec{r})}[\phi(t)],
   \label{eq:mudef}
\end{align}
which leads to a local force $-\nabla\mu(\vec{r},[\phi(t)])$ that generates a
flux \cite{Dieterich1990}
\begin{align}
   \vec{j}(\vec{r},[\phi(t)]) = 
   -\frac{D\rho_\text{max}}{k_BT}\phi(\vec{r},t)\nabla\mu(\vec{r},[\phi(t)])
   \label{eq:jdef}
\end{align}
with the collective diffusion constant $D$.
From the continuity equation of the particle number one obtains the conserved
dynamics (model B \cite{Hohenberg1977}) equation of motion of 
$\phi(\vec{r},t)$:
\begin{align}
   \rho_\text{max}\frac{\partial}{\partial t}\phi(\vec{r},t) = 
   -\nabla\cdot\vec{j}(\vec{r},[\phi(t)]).
   \label{eq:conti}
\end{align}

In the low-density limit, $\phi(\vec{r},t)\ll1$, Eqs.~\Eq{mudef}--\Eq{conti}
lead to the well-known diffusion equation $\partial\phi(\vec{r},t)/\partial t = 
D\nabla^2\phi(\vec{r},t)$.
However, in general, Eqs.~\Eq{mudef}--\Eq{conti} do \emph{not} correspond to a
simple diffusion equation of the occupancy $\phi(\vec{r},t)$ for at least three
reasons:
First, at higher densities the effective diffusion ``constant'' becomes
$\phi(\vec{r},t)$-dependent, which renders the problem non-linear. 
Second, upon deriving an approximate diffusion equation from 
Eqs.~\Eq{mudef}--\Eq{conti} one has to decide around which reference density to
expand, and there is no unique natural choice in the context of liquid-liquid 
interfaces.
Third, the driving force for the diffusion process described by 
Eqs.~\Eq{mudef}--\Eq{conti} is \emph{not} a non-vanishing gradient in the 
\emph{density} (consider, e.g., a liquid-liquid interface at equilibrium), but
in the \emph{local chemical potential} (see Eq.~\Eq{jdef}).
Hence any derivation of a diffusion equation introduces some \emph{assumption} 
concerning the final structure at the end of the relaxation process.
However, Eqs.~\Eq{mudef}--\Eq{conti} lead uniquely, i.e., without further 
assumptions, from any initial state to that final state which is consistent
with the underlying density functional. 

In the following a planar surface between liquid and vapor bulk states (see
Fig.~\ref{fig:pd}) will be considered.
The lateral translational symmetry leads to profiles 
$\phi(z,t)=\widetilde{\phi}(\widetilde{z},\widetilde{t})$ which depend only on
the coordinate $z=\widetilde{z}\lambda$ normal to the surface and on the time 
$t=\widetilde{t}\tau$ with the diffusion time $\tau=\lambda^2/D$. 
Using Eqs.~\Eq{U}--\Eq{Frpa} one obtains from Eq.~\Eq{mudef}
\begin{align}
   \widetilde{\mu}(\widetilde{z}, & [\widetilde{\phi}])
   := \frac{\mu(z,[\phi])}{k_BT} \notag\\
   &= \ln\left(\frac{\widetilde{\phi}(\widetilde{z})}
                    {1-\widetilde{\phi}(\widetilde{z})}\right)
      - \frac{3}{2T^*}\int\limits_{-1}^1\!\d u\;(1-u^2)
        \widetilde{\phi}(\widetilde{z}+u). 
   \label{eq:mu1d} 
\end{align}
Equation~\Eq{jdef} leads to
\begin{align}
   \widetilde{j}(\widetilde{z},[\widetilde{\phi}])
   := \frac{j(z,[\phi])}{\rho_\text{max}\lambda/\tau} 
   = -\widetilde{\phi}(\widetilde{z})
     \frac{\partial}{\partial \widetilde{z}}
     \widetilde{\mu}(\widetilde{z},[\widetilde{\phi}])
   \label{eq:j1d}
\end{align}
and Eq.~\Eq{conti} takes the form
\begin{align}
   \frac{\partial}{\partial \widetilde{t}}
   \widetilde{\phi}(\widetilde{z},\widetilde{t}) 
   = -\frac{\partial}{\partial \widetilde{z}}
     \widetilde{j}(\widetilde{z},[\widetilde{\phi}(\widetilde{t})]).
   \label{eq:conti1d}
\end{align}
Equations~\Eq{mu1d}--\Eq{conti1d} demonstrate that, when expressed in terms of
the interaction range $\lambda$ and the diffusion time $\tau$, the temporal 
evolution of the state of the system 
$\widetilde{\phi}(\widetilde{z},\widetilde{t})$ depends only on one parameter:
the effective temperature $T^*$.

In the present study Eqs.~\Eq{mu1d}--\Eq{conti1d} are solved numerically 
subject to the boundary conditions 
\begin{align}
   \widetilde{\phi}(\widetilde{z}\to-\infty,\widetilde{t}>0)\to\phi_\text{L}, 
   \label{eq:bcL} \\
   \widetilde{\phi}(\widetilde{z}\to+\infty,\widetilde{t}>0)\to\phi_\text{V}, 
   \label{eq:bcV}
\end{align}
where the bulk occupancies $\phi_\text{L}$ and $\phi_\text{V}$ correspond to
the reduced bulk chemical potential differences from coexistence 
$\Delta\widetilde{\mu}_\text{L}:=\widetilde{\mu}_\text{L}-
\widetilde{\mu}_\text{co}(T^*)\geq0$ (liquid) and 
$\Delta\widetilde{\mu}_\text{V}:=\widetilde{\mu}_\text{V}-
\widetilde{\mu}_\text{co}(T^*)\leq0$ (vapor), respectively, at a given 
temperature $T^*\leq T^*_\text{c}$ (see Fig.~\ref{fig:pd}).

Since our interest here is not in describing a realistic system quantitatively
but in inferring the general mechanism of interface formation, we consider
the artificial but simple initial state 
\begin{align}
   \widetilde{\phi}(\widetilde{z},\widetilde{t}=0) =
   \begin{cases}
      \phi_\text{L} & \text{, $\widetilde{z} <    0$}    \\
      \phi_\text{V} & \text{, $\widetilde{z} \geq 0$}
   \end{cases}.   
   \label{eq:init}
\end{align}
The conclusions to be drawn later are not expected to depend on this choice. 


\subsection{Gibbs dividing surface}

\begin{figure*}[!t]
   \includegraphics{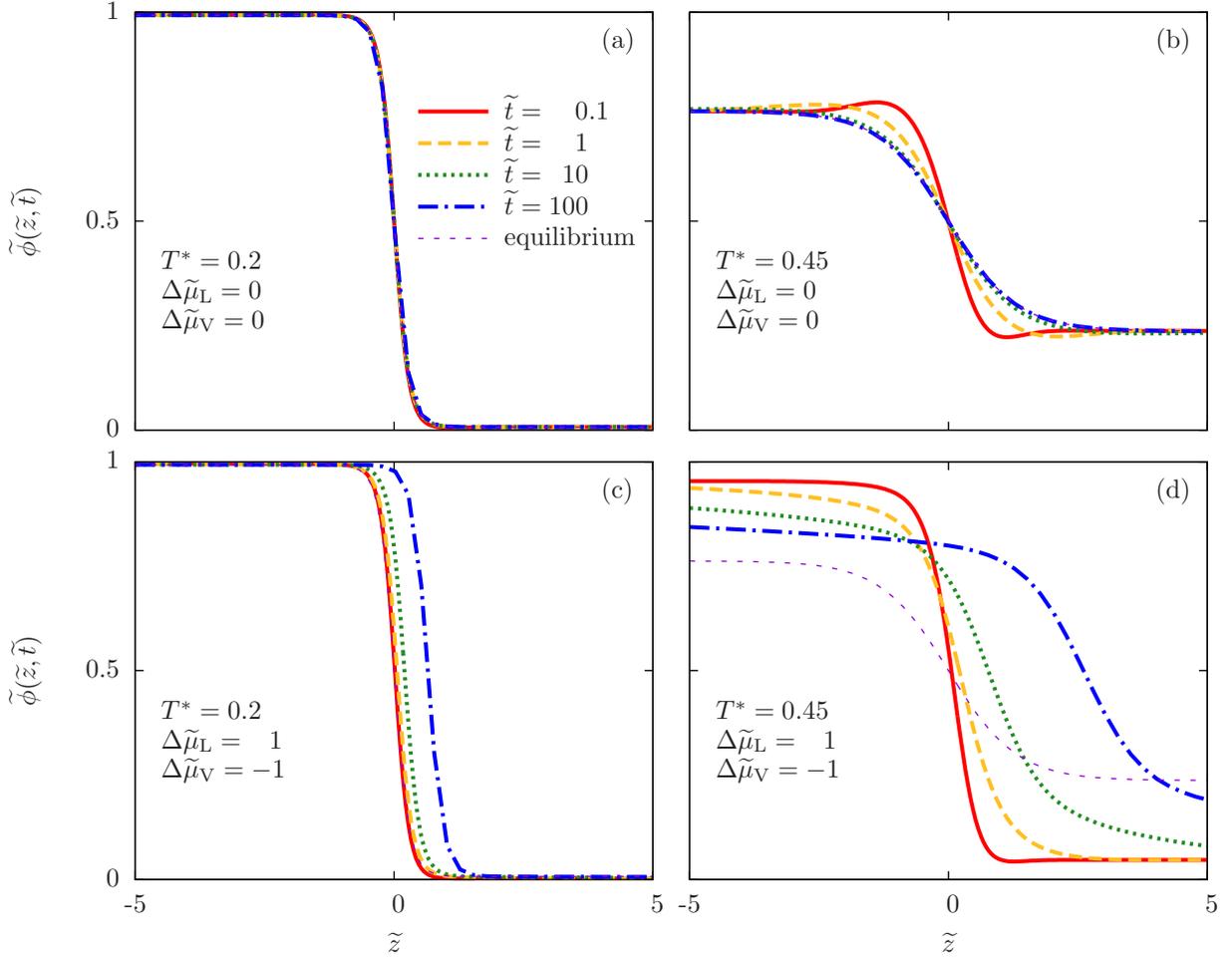}
   \caption{Temporal evolution of the occupancy profiles 
           $\widetilde{\phi}(\widetilde{z},\widetilde{t})$ in terms of the 
           dimensionless position $\widetilde{z}=z/\lambda$ and the 
           dimensionless time $\widetilde{t}=t/\tau$ for reduced temperatures
           $T^*$ and reduced bulk chemical potential differences from
           coexistence $\Delta\widetilde{\mu}_\text{L,V}$.
           Panels (a) and (b) display the relaxation towards the equilibrium 
           interfaces (thin dashed lines) between bulk phases at coexistence,
           which slows down upon approaching the critical temperature
           $T^*_\text{c}=1/2$.
           The symmetry with respect to the critical occupancy $\phi_\text{c}
           =1/2$ is an irrelevant artifact of the lattice-gas expression 
           Eq.~\Eq{Fhs}.
           Panels (c) and (d) exemplify the relaxation and translation
           of the interface in the absence of two-phase coexistence.
           Far below the critical temperature $T^*_\text{c}$ (see panel (c))
           the interface relaxation is faster than the translation, whereas 
           the opposite occurs close to the critical point (see panel (d)).
           For long times during the early-stage relaxation process the 
           structures of the non-equilibrium interfaces become identical to
           the equilibrium interfaces at the corresponding temperatures $T^*$.}
   \label{fig:profiles}
\end{figure*}

In order to quantify the amount of colloidal particles evaporating from the 
liquid into the vapor phase the position of the Gibbs dividing surface 
$z_\text{GDS}(t)$ at time $t$ is introduced by \cite{Rowlinson2002}
\begin{align}
   \int\limits_{-\infty}^{z_\text{GDS}(t)}\!\d z\;(\phi(z,t)-\phi_\text{L}) +
   \int\limits_{z_\text{GDS}(t)}^\infty\!\d z\;(\phi(z,t)-\phi_\text{V}) = 0.
   \label{eq:zGDS}
\end{align}
In the light of the continuously spreading tails of the liquid-liquid interface
to be observed later (see, e.g., Fig.~\ref{fig:profiles}(d)) the notion of a
Gibbs dividing surface might appear somewhat unusual, but it is only the 
aspect of the \emph{position} and not that of the (diverging) width which is 
used here.
Taking the derivative of the previous equation with respect to time 
$t$ leads to 
\begin{align}
   (\phi_\text{L}-\phi_\text{V})\frac{\d z_\text{GDS}(t)}{\d t}
   &= \int\!\d z\;\frac{\partial \phi(z,t)}{\partial t} 
   \notag\\
   &= -\frac{1}{\rho_\text{max}}\int\!\d z\;\frac{\partial j(z,t)}{\partial z}
   \notag\\
   &= -\frac{1}{\rho_\text{max}}(j(\infty,t)-j(-\infty,t))
   \notag\\
   &= 0. 
\end{align}
Therefore, the position of the Gibbs dividing surface is independent of time:
$z_\text{GDS}(t)=z_\text{GDS}(0)=0$ (see Eq.~\Eq{init}).
Consequently, the excess number of particles per cross-sectional area in the
liquid phase
\begin{align}
   \Gamma_\text{L}(t) := 
   \rho_\text{max}\int\limits_{-\infty}^0\!\d z\;
   (\phi(z,t)-\phi_\text{L})
   \label{eq:GammaL}
\end{align}
changes with the rate
\begin{align}
   \frac{\d \Gamma_\text{L}(t)}{\d t}
   &= \rho_\text{max}\int\limits_{-\infty}^0\!\d z\;
      \frac{\partial \phi(z,t)}{\partial t} 
    = -\int\limits_{-\infty}^0\!\d z\;
      \frac{\partial j(z,t)}{\partial z}
   \notag\\
   &= -(j(0,t)-j(-\infty,t))
    = -j(0,t). 
\end{align}
This finding is to be expected due to the underlying conserved dynamics of the
colloidal fluid.
A dimensionless form of Eq.~\Eq{GammaL} is given by
\begin{align}
   \widetilde{\Gamma}_\text{L}(\widetilde{t}) 
   := \frac{\Gamma_\text{L}(t)}{\rho_\text{max}\lambda}
   = \int\limits_{-\infty}^0\!\d \widetilde{z}\;
     (\widetilde{\phi}(\widetilde{z},\widetilde{t})-\phi_\text{L}).
   \label{eq:GammaL2}
\end{align}


\subsection{Interfacial tension}

The interfacial tension is defined as the work per interfacial area which is 
required to increase the interfacial area $A$ but keeping the total volume $V$ 
as well as the number of particles $N$ constant \cite{Rowlinson2002}:
\begin{align}
   \gamma[\phi] = \left(\frac{\partial F[\phi]}{\partial A}\right)_{V,N}.
   \label{eq:gamma1}
\end{align}
In the present case of a colloidal dispersion a change of the interfacial area
between the colloidal liquid and the colloidal vapor could be achieved by
deforming the container of the fluid appropriately, which leads to a 
deformation of the incompressible molecular solvent dragging the dispersed
colloids with it.
Due to the separation of molecular and colloidal time scales, one is able to
perform this deformation, on the one hand, sufficiently slow in order to stay
in the regime of low Reynolds numbers to avoid dissipation due to turbulence,
and, on the other hand, sufficiently fast such that the colloidal distribution
$\phi(z,t)$ is practically not evolving during the measurement.

Consider the system volume $\mathcal{V}\subseteq\mathbb{R}^3$ and a
divergence-free map $\vec{w}: \mathcal{V}\to\mathbb{R}^3$, i.e.,
$\div\vec{w}=0$, which corresponds to the incompressibility of the solvent. 
This leads to a shift of any point $\vec{r}\in\mathcal{V}$ to the new position
$\vec{r}_\vec{w}:=\vec{r}+\vec{w}(\vec{r})$ as well as to the deformation of 
the system volume $\mathcal{V}$ into $\mathcal{V}_\vec{w} :=
\{\vec{r}_\vec{w}\in\mathbb{R}^3 | \vec{r}\in\mathcal{V}\}$.
Moreover, $\div\vec{w}=0$ implies that the number density of colloids does
not change due to deformation $\vec{w}$: $\phi_\vec{w}(\vec{r}_\vec{w}) =
\phi(\vec{r})$.

The free energy $F_\vec{w}[\phi_\vec{w}]$ of the deformed system 
$\mathcal{V}_\vec{w}$ in state $\phi_\vec{w}$ is given by (see 
Eqs.~\Eq{df3d}--\Eq{Frpa})
\begin{widetext}
\begin{align}
   F_\vec{w}[\phi_\vec{w}]
   &=  
   k_BT\rho_\text{max}\int\limits_{\mathcal{V}_\vec{w}}\!\d^3r_\vec{w}\;
   \big[\phi_\vec{w}(\vec{r}_\vec{w}) 
        \ln(\phi_\vec{w}(\vec{r}_\vec{w})) +
        (1-\phi_\vec{w}(\vec{r}_\vec{w}))
        \ln(1-\phi_\vec{w}(\vec{r}_\vec{w}))\big] 
   \notag\\   
   &\phantom{=}+
   \frac{\rho_\text{max}^2}{2}
   \int\limits_{\mathcal{V}_\vec{w}}\!\d^3r_\vec{w}\!
   \int\limits_{\mathcal{V}_\vec{w}}\!\d^3r'_\vec{w}\;
   U(|\vec{r}_\vec{w}-\vec{r'}_\vec{w}|)
   \phi_\vec{w}(\vec{r}_\vec{w})\phi_\vec{w}(\vec{r'}_\vec{w}) 
   \notag\\
   &=  
   k_BT\rho_\text{max}\int\limits_{\mathcal{V}}\!\d^3r\;
   \big[\phi(\vec{r})
        \ln(\phi(\vec{r})) +
        (1-\phi(\vec{r}))
        \ln(1-\phi(\vec{r}))\big] 
   \notag\\   
   &\phantom{=}+ 
   \frac{\rho_\text{max}^2}{2}
   \int\limits_{\mathcal{V}}\!\d^3r\!
   \int\limits_{\mathcal{V}}\!\d^3r'\;
   U(|\vec{r}-\vec{r'}+\vec{w}(\vec{r})-\vec{w}(\vec{r'})|)
   \phi(\vec{r})\phi(\vec{r'}).
   \label{eq:Fw}
\end{align}
\end{widetext}
Obviously only the RPA part of $F_\vec{w}[\phi_\vec{w}]$ depends on
the the deformation $\vec{w}$, i.e., only this double volume integral
contributes to the interfacial tension Eq.~\Eq{gamma1}.

If the deformation $\vec{w}$ stretches the Cartesian $z$-component $z-z'$ of 
the difference vector $\vec{r}-\vec{r'}$ between two points 
$\vec{r},\vec{r'}\in\mathcal{V}$ by a factor $\eta$ and the projection 
$\Delta\vec{r}_\perp$ onto the $x$-$y$-plane by, due to $\div\vec{w}=0$, a 
factor $-\eta/2$, i.e.,
\begin{align}
    & |\vec{r}-\vec{r'}+\vec{w}(\vec{r})-\vec{w}(\vec{r'})| \notag\\
    & =\sqrt{(1-\eta/2)^2\Delta r_\perp^2 + (1+\eta)^2(z-z')^2},
\end{align} 
the RPA part of $F_\vec{w}[\phi_\vec{w}]$ equals 
\begin{align}
   -\frac{\pi\epsilon\rho_\text{max}^2A}{2}
   \int\!\d z\;\phi(z)
   \!\!\!\!\!\!\!\!\!\!
   \int\limits_{z-\lambda/(1+\eta)}^{z+\lambda/(1+\eta)}
   \!\!\!\!\!\!\!\!\!\!
   \d z'\;\frac{\lambda^2-(1+\eta)^2(z-z')^2}{(1-\eta/2)^2}\phi(z')
\end{align}
and the cross-sectional area $A$ is deformed to $A_\vec{w}=(1-\eta/2)^2A$.

Equation~\Eq{gamma1} leads to
\begin{align}
   \gamma[\phi] 
   &= \left.
      \frac{\dps\d F_\vec{w}[\phi_\vec{w}]/\d \eta}
           {\dps\d A_\vec{w}/\d \eta}	  
      \right|_{\eta=0} \notag\\
   &= \frac{\pi\epsilon\rho_\text{max}^2}{2}
      \int\!\d z\;\phi(z)
      \!\!
      \int\limits_{z-\lambda}^{z+\lambda}
      \!\!
      \d z'\;(\lambda^2-3(z-z')^2)\phi(z') 
      \label{eq:gamma2}
\end{align}   
and with Eq.~\Eq{tstar} to
\begin{align}
   \widetilde{\gamma}[\widetilde{\phi}]
   &:= \frac{\gamma[\phi]}{k_BT\rho_\text{max}\lambda}
   \notag\\
   &\phantom{:}= 
      \frac{3}{4T^*}
      \int\!\d \widetilde{z}\;\widetilde{\phi}(\widetilde{z})
      \int\limits_{-1}^1
      \!\d u\;(1-3u^2)\widetilde{\phi}(\widetilde{z}+u). 
      \label{eq:gamma3}
\end{align}


\subsection{Dissipation rate}

The rate of energy dissipation per cross-sectional area is denoted by
\begin{align}
   P(t) 
   &= -\frac{\d F[\phi(t)]/A}{\d t}
   \notag\\
   &\phantom{:}= 
      -\frac{1}{A}\int\!\d^3r\;\frac{\delta F}{\delta\phi(\vec{r})}[\phi(t)]
      \frac{\partial\phi(\vec{r},t)}{\partial t}
   \notag\\
   &\phantom{:}= 
      \frac{1}{A}\int\!\d^3r\;
      \mu(\vec{r},[\phi(t)])\nabla\cdot\vec{j}(\vec{r},[\phi(t)]).
\end{align}
Since the normal component of the flux $\vec{j}$ vanishes at the system 
boundaries, an integration by parts using the Gaussian integral theorem leads
to 
\begin{align}
   P(t)
   &= \frac{1}{A}\int\!\d^3r\;
      (-\nabla\mu(\vec{r},[\phi(t)]))\cdot\vec{j}(\vec{r},[\phi(t)])
   \notag\\
   &= \frac{k_BT}{D\rho_\text{max}A}\int\!\d^3r\;
      \frac{\vec{j}(\vec{r},[\phi(t)])^2}{\phi(\vec{r},t)}
\end{align}
and hence
\begin{align}
   \widetilde{P}(\widetilde{t})
   := \frac{P(t)}{k_BT\rho_\text{max}\lambda/\tau}
   = \int\!\d \widetilde{z}\;
     \frac{\widetilde{j}(\widetilde{z},[\widetilde{\phi}(\widetilde{t})])^2}
          {\widetilde{\phi}(\widetilde{z},\widetilde{t})}.
   \label{eq:P}
\end{align}


\begin{figure*}[!t]
   \includegraphics{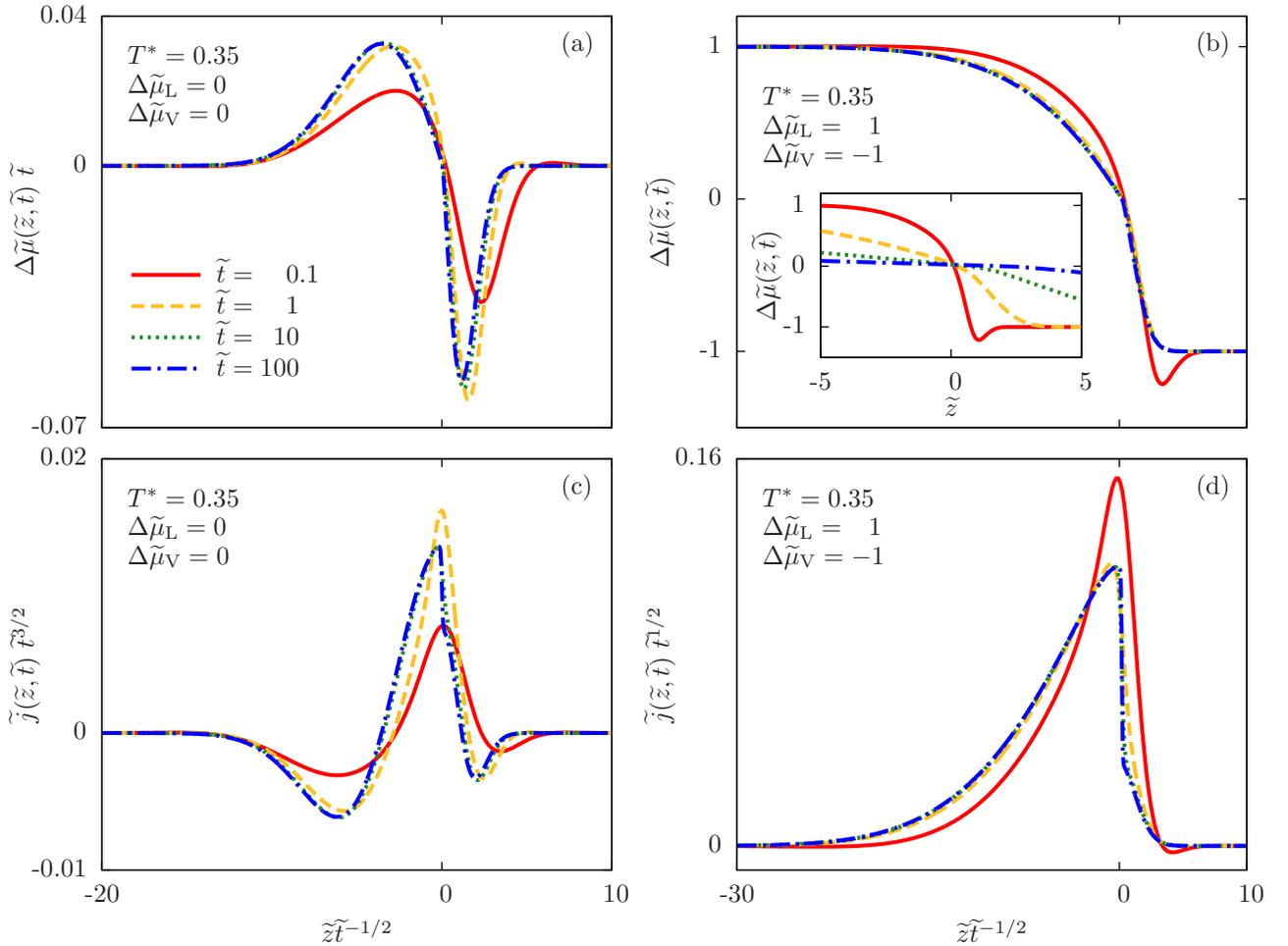}
   \caption{Scaling behavior of the reduced local chemical potential
           difference from coexistence 
           $\Delta\widetilde{\mu}(\widetilde{z},\widetilde{t})$ and of the 
           reduced flux $\widetilde{j}(\widetilde{z},\widetilde{t})$ in terms 
           of the dimensionless position $\widetilde{z}=z/\lambda$ and the 
           dimensionless time $\widetilde{t}=t/\tau$ for reduced temperature
           $T^*=0.35$.
           For both bulk phases at coexistence the leading contributions
           are of the forms 
           $\Delta\widetilde{\mu}(\widetilde{z},\widetilde{t}\to\infty)
           \simeq \widetilde{t}^{-1}M_2(\widetilde{z}\widetilde{t}^{-1/2})$
           (panel (a)) and
           $\widetilde{j}(\widetilde{z},\widetilde{t}\to\infty)
           \simeq \widetilde{t}^{-3/2}J_2(\widetilde{z}\widetilde{t}^{-1/2})$
           (panel (c)), whereas in the absence of two-phase coexistence the 
           leading contributions
           are given by
           $\Delta\widetilde{\mu}(\widetilde{z},\widetilde{t}\to\infty)
           \simeq M_1(\widetilde{z}\widetilde{t}^{-1/2})$
           (panel (b)) and
           $\widetilde{j}(\widetilde{z},\widetilde{t}\to\infty)
           \simeq \widetilde{t}^{-1/2}J_1(\widetilde{z}\widetilde{t}^{-1/2})$
           (panel (d)).
           The inset in panel (b) illustrates the temporal increase
           of the spatial range with $\widetilde{\mu}(\widetilde{z},
           \widetilde{t})\approx\widetilde{\mu}_\text{co}(T^*)$ around the 
           Gibbs dividing interface at $\widetilde{z}=0$.}
   \label{fig:muj}
\end{figure*}

\section{\label{sec:results}Results and Discussion}

\subsection{\label{subsec:structure}Interfacial structure}

Numerical solutions $\widetilde{\phi}(\widetilde{z},\widetilde{t})$ of 
Eqs.~\Eq{mu1d}--\Eq{bcV} are displayed in Fig.~\ref{fig:profiles} for the two
representative reduced temperatures $T^*=0.2$ (panels (a) and (c)) and 
$T^*=0.45$ (panels (b) and (d)). 
The latter temperature is close to the reduced critical temperature 
$T^*_\text{c}=1/2$, while the former temperature corresponds to a typical 
triple point temperature, which for many fluids is approximately $40\%$ of the
critical temperature \cite{Lide1998}.

Figures~\ref{fig:profiles}(a) and (b) correspond to the relaxation towards
the equilibrium interface between both bulk phases at coexistence 
($\Delta\widetilde{\mu}_\text{L}=\Delta\widetilde{\mu}_\text{V}=0$).
At low reduce temperatures $T^*$ the interface forms rapidly (see 
Fig.~\ref{fig:profiles}(a)), whereas the interface formation is slowed
down close to the critical point (see Fig.~\ref{fig:profiles}(b)).

Non-equilibrium conditions are exemplified in Figs.~\ref{fig:profiles}(c) 
and (d) by an oversaturated colloidal liquid 
($\Delta\widetilde{\mu}_\text{L}=1>0$) in contact with an
undersaturated colloidal vapor ($\Delta\widetilde{\mu}_\text{V}=
-1<0$).
Besides the formation of the liquid-vapor interface a drift due to the
chemical potential difference between the bulk phases is observed.
At low reduce temperatures $T^*$ the interface forms before a significant
drift occurs (see Fig.~\ref{fig:profiles}(c)), whereas close to the critical
point the interface formation is slowed down such that the interface drift
sets in before (see Fig.~\ref{fig:profiles}(d)).

The remarkable observation to be made in Figs.~\ref{fig:profiles}(c) and (d)
is that even for strong non-equilibrium conditions 
($|\Delta\widetilde{\mu}_\text{L,V}|\not\ll1$) the interfacial structures 
at become identical to that of the equilibrium interfaces (thin dashed lines)
at the respective reduced temperature $T^*$.

In order to understand the formation process of interfaces, the reduced local
chemical potential difference from coexistence 
$\Delta\widetilde{\mu}(\widetilde{z},\widetilde{t})=
\widetilde{\mu}(\widetilde{z},\widetilde{t})-\widetilde{\mu}_\text{co}(T^*)$
and the reduced flux $\widetilde{j}(\widetilde{z},\widetilde{t})$ are displayed
in Fig.~\ref{fig:muj} for the intermediate reduced temperature $T^*=0.35$.
In the case of both bulk phases being at coexistence 
($\Delta\widetilde{\mu}_\text{L,V}=0$) one infers a scaling behavior
$\Delta\widetilde{\mu}(\widetilde{z},\widetilde{t}\to\infty)
\simeq \widetilde{t}^{-1}M_2(\widetilde{z}\widetilde{t}^{-1/2})$
(Fig.~\ref{fig:muj}(a)) and
$\widetilde{j}(\widetilde{z},\widetilde{t}\to\infty)
\simeq \widetilde{t}^{-3/2}J_2(\widetilde{z}\widetilde{t}^{-1/2})$
(Fig.~\ref{fig:muj}(c)) with scaling functions $M_2(x)$ and $J_2(x)$.
However, in the absence of two-phase coexistence 
$\Delta\widetilde{\mu}(\widetilde{z},\widetilde{t}\to\infty)
\simeq M_1(\widetilde{z}\widetilde{t}^{-1/2})$ (Fig.~\ref{fig:muj}(b)) and
$\widetilde{j}(\widetilde{z},\widetilde{t}\to\infty)
\simeq \widetilde{t}^{-1/2}J_1(\widetilde{z}\widetilde{t}^{-1/2})$
(Fig.~\ref{fig:muj}(d)) with scaling functions $M_1(x)$ and $J_1(x)$ 
is found.
The scaling functions $M_{1,2}(x)$ and $J_{1,2}(x)$ in general depend on 
$T^*$, $\Delta\widetilde{\mu}_\text{L}$, and $\Delta\widetilde{\mu}_\text{V}$.
The total scaling behavior may be combined to
\begin{align}
   \Delta\widetilde{\mu}(\widetilde{z},\widetilde{t}\to\infty)
   \simeq 
   M_1(\widetilde{z}\widetilde{t}^{-1/2}) +
   \widetilde{t}^{-1}M_2(\widetilde{z}\widetilde{t}^{-1/2})
   \label{eq:muscaling}
\end{align}
and
\begin{align}
   \widetilde{j}(\widetilde{z},\widetilde{t}\to\infty)
   \simeq 
   \widetilde{t}^{-1/2}J_1(\widetilde{z}\widetilde{t}^{-1/2}) +
   \widetilde{t}^{-3/2}J_2(\widetilde{z}\widetilde{t}^{-1/2}).
   \label{eq:jscaling}
\end{align}
Obviously, Eq.~\Eq{jscaling} follows from Eq.~\Eq{muscaling} by means of
Eq.~\Eq{j1d}.
For two-phase coexistence the scaling function $M_1(x)$ in Eq.~\Eq{muscaling}
vanishes so that the leading contribution is given by
the second term on the right-hand side of Eq.~\Eq{muscaling}.
In the absence of two-phase coexistence the scaling function $M_1(x)$ does
not vanish identically so that it gives rise to the leading order contribution.

Considering the situation of a non-equilibrium interface displayed in 
Fig.~\ref{fig:muj}(b), one observes that the reduced local chemical potential
$\widetilde{\mu}(\widetilde{z}\approx0,\widetilde{t})$ in the vicinity of the 
Gibbs dividing interface is close to the coexistence value 
$\widetilde{\mu}_\text{co}(T^*)=-1/T^*$.
Then the scaling  $\Delta\widetilde{\mu}(\widetilde{z},\widetilde{t}\to\infty)
\simeq M_1(\widetilde{z}\widetilde{t}^{-1/2})$ implies 
$\widetilde{\mu}(\widetilde{z},\widetilde{t}\to\infty)\approx
\widetilde{\mu}_\text{co}(T^*)$ for an increasingly wide range 
$|\widetilde{z}|<\widetilde{R}(\widetilde{t})\sim\widetilde{t}^{1/2}$ around 
the Gibbs dividing interface (see inset in Fig.~\ref{fig:muj}(b)).
Once $\widetilde{R}(\widetilde{t})$ exceeds the bulk correlation length, the
local chemical potential in the interfacial range coincides with that of the
equilibrium interface.
Therefore, for during the early-stage relaxation process non-equilibrium 
interfaces approach the same structure as the equilibrium interface at the same
temperature, because the finite chemical potential difference between the bulk
phases is bridged by the local chemical potential over an unlimited spatial
range so that locally, at the interface position, an effectively uniform
chemical potential profile is sensed by the fluid.


\subsection{Interfacial tension}

\begin{figure}[!t]
   \includegraphics{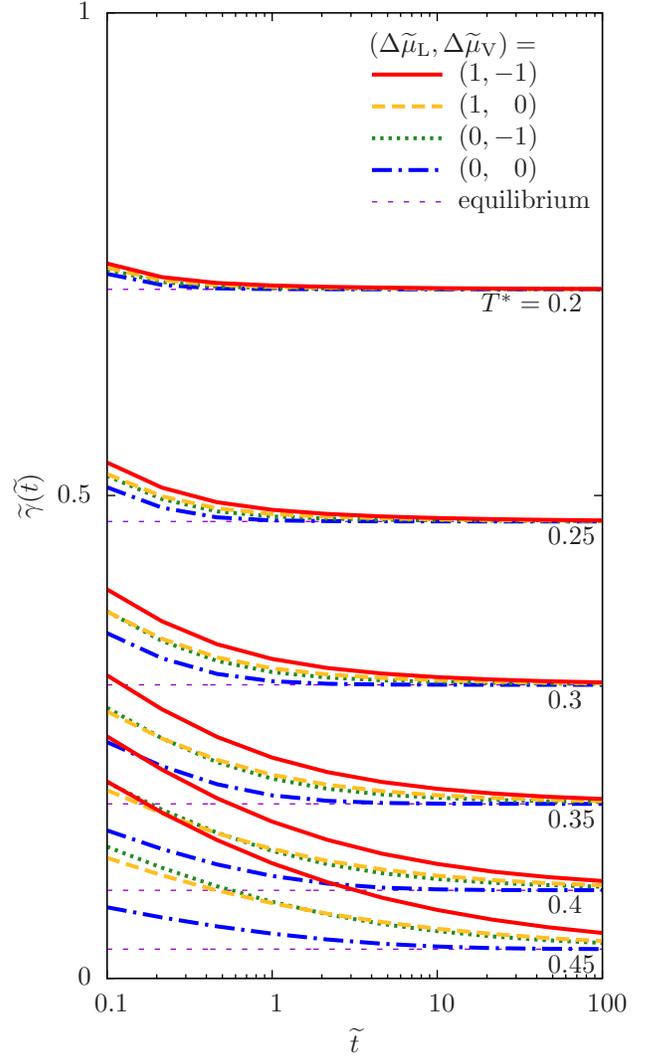}
   \caption{Temporal evolution of the reduced interfacial tension
           $\widetilde{\gamma}(\widetilde{t})$ in terms of the dimensionless
           time $\widetilde{t}=t/\tau$ for various reduced temperatures
           $T^*$. 
           Irrespective of the states $\Delta\widetilde{\mu}_\text{L,V}$ of
           the bulk phases, the interfacial tension approaches
           the values of the equilibrium interfaces (thin horizontal dashed 
           lines) at the respective reduced temperatures $T^*$.}
   \label{fig:gammalin}
\end{figure}

The phenomenon of non-equilibrium interfaces to approach the structure of 
equilibrium interfaces discussed in the previous 
Subsec.~\ref{subsec:structure} implies the approach of the interfacial 
tension Eq.~\Eq{gamma3} towards the equilibrium value.
This expectation is confirmed in Fig.~\ref{fig:gammalin}, which displays the
interfacial tension $\widetilde{\gamma}(\widetilde{t})=
\widetilde{\gamma}[\widetilde{\phi}(\widetilde{t})]$ as a function of the
dimensionless time $\widetilde{t}$ for various equilibrium and non-equilibrium
conditions of the bulk phases and several reduced temperatures $T^*$.
The curves approach the equilibrium values (thin horizontal dashed lines) at 
the respective temperatures.
As is intuitively expected the initial deviations of the interfacial
tension from its equilibrium value increase with the deviations of the two bulk
fluids from two-phase coexistence.
Consequently the larger the deviation from two-phase coexistence is, the longer
the relaxation of the interfacial tension to its final equilibrium value
takes. 

\begin{figure}[!t]
   \includegraphics{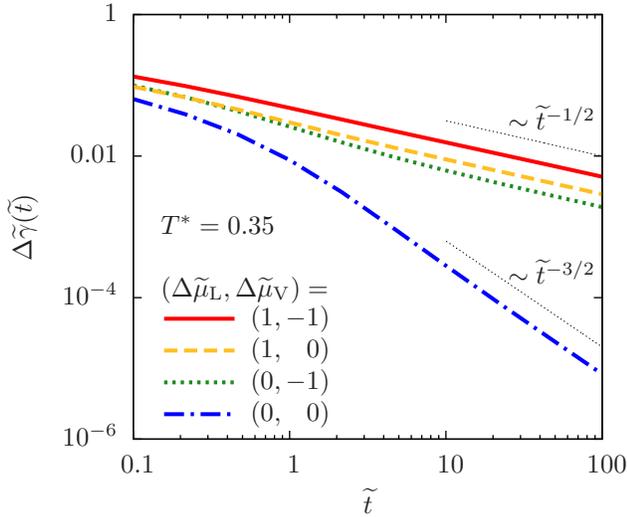}
   \caption{Deviation $\Delta\widetilde{\gamma}(\widetilde{t})$ of the
           interfacial tension from the equilibrium value as a function of the
           dimensionless time $\widetilde{t}=t/\tau$ for reduced temperature
           $T^*=0.35$.
           As a function of time $\widetilde{t}$ the relaxation occurs 
           algebraically according to $\Delta\widetilde{\gamma}(\widetilde{t}
           \to\infty)\sim\widetilde{t}^{-3/2}$ for both bulk phases at 
           coexistence and $\Delta\widetilde{\gamma}(\widetilde{t}\to\infty)
           \sim\widetilde{t}^{-1/2}$ for non-equilibrium conditions.}
   \label{fig:gammalog}
\end{figure}

The actual mode of relaxation towards the equilibrium value of the interfacial
tension can be inferred from Fig.~\ref{fig:gammalog}, which displays the 
difference $\Delta\widetilde{\gamma}(\widetilde{t})$ of the interfacial tension
at dimensionless time $\widetilde{t}$ from the final equilibrium value
for the reduced temperature $T^*=0.35$.
It can be observed that for two-phase coexistence the relaxation occurs 
according to $\Delta\widetilde{\gamma}(\widetilde{t})\sim \widetilde{t}^{-3/2}$,
whereas in the absence of two-phase coexistence the decay is of the form
$\Delta\widetilde{\gamma}(\widetilde{t})\sim \widetilde{t}^{-1/2}$.
The algebraic relaxation of the interfacial tension is related to 
the underlying conserved dynamics, which requires rearrangements of the
fluid to occur by transport.
The observation that the relaxation in the case of two-phase coexistence is
faster (with exponent $-3/2$) than in the absence of two-phase coexistence
(with exponent $-1/2$) can be understood by noting that in the former situation 
only the interface relaxes, whereas in the latter situation the interface
relaxation occurs on top of a diffusive flow from one bulk phase to the other.


\subsection{Energy dissipation}

\begin{figure}[!t]
   \includegraphics{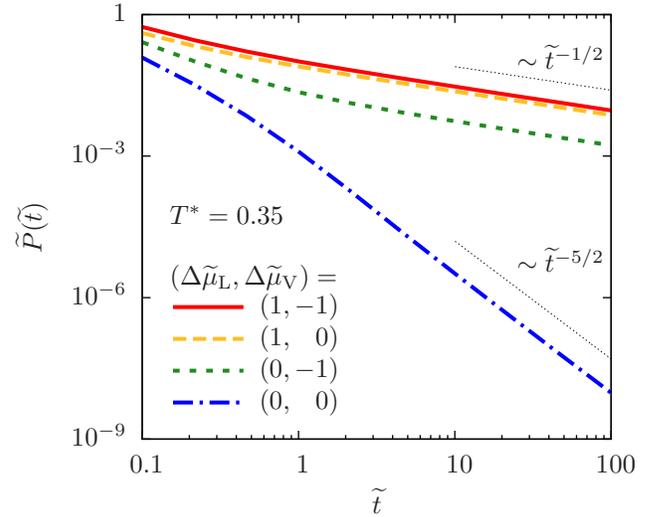}
   \caption{Reduced energy dissipation rate $\widetilde{P}(\widetilde{t})$ per
           cross-sectional area as a function of the dimensionless time 
           $\widetilde{t}=t/\tau$ for reduced temperature $T^*=0.35$.
           As a function of time $\widetilde{t}$ the dissipation decays 
           algebraically according to $\widetilde{P}(\widetilde{t}\to\infty)
           \sim\widetilde{t}^{-5/2}$ for both bulk phases at coexistence and
           $\widetilde{P}(\widetilde{t}\to\infty)\sim
           \widetilde{t}^{-1/2}$ for non-equilibrium conditions.}
   \label{fig:P}
\end{figure}

During the formation of the interface or the evaporation of particles from the
oversaturated into the undersaturated bulk phase energy is dissipated, which
corresponds to a decrease of the total Helmholtz free energy $F[\phi(t)]$.
Since an isothermal system is considered here the dissipated energy is 
absorbed by the heat bath.
Figure~\ref{fig:P} displays the reduced energy dissipation rate 
$\widetilde{P}(\widetilde{t})$ defined in Eq.~\Eq{P} for reduced temperature
$T^*=0.35$ and various equilibrium and non-equilibrium conditions.
The dissipation rate decays algebraically
according to $\widetilde{P}(\widetilde{t}\to\infty)\sim\widetilde{t}^{-5/2}$
for two-phase coexistence and $\widetilde{P}(\widetilde{t}\to\infty)\sim
\widetilde{t}^{-1/2}$ otherwise.
These exponents are a simple consequence of Eq.~\Eq{jscaling} used in 
Eq.~\Eq{P}.

The origin of the observed differences between the case of two-phase 
coexistence and that of non-equilibrium conditions is again the fact that in
the absence of two-phase coexistence dissipation is dominated by the diffusive
flow induced by the two bulk phases being not at equilibrium with each other,
which does not occur for two-phase coexistence.


\subsection{Evaporation}

\begin{figure}[!t]
   \includegraphics{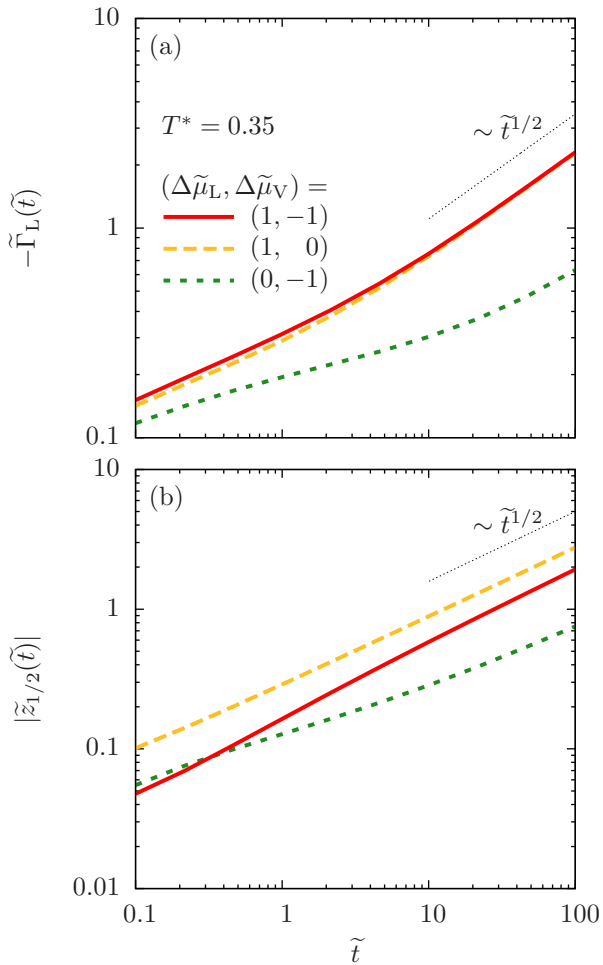}
   \caption{Reduced number of evaporated particles 
           $-\widetilde{\Gamma}_\text{L}(\widetilde{t})$ per cross-sectional 
           area (panel (a)) and absolute reduced interface position
           $|\widetilde{z}_{1/2}(\widetilde{t})|$, where 
           $\widetilde{\phi}(\widetilde{z}_{1/2}(\widetilde{t}),\widetilde{t})
           =1/2$, (panel (b)) as functions of the reduced time 
           $\widetilde{t}=t/\tau$ for reduced temperature $T^*=0.35$.
           Both quantities increase algebraically $\sim\widetilde{t}^{1/2}$.}
   \label{fig:evaporation}
\end{figure}

In the absence of two-phase coexistence with an oversaturated liquid and/or
an undersaturated vapor phase one expects evaporation of the liquid.
This terminology is used here also in the context of colloidal
fluids with colloid-rich (liquid) and colloid-poor (vapor) phases.
Amongst the various possibilities to quantify evaporation two are shown in 
Fig.~\ref{fig:evaporation}.

Figure~\ref{fig:evaporation}(a) displays the reduced number of evaporated
particles per cross-sectional area 
$-\widetilde{\Gamma}_\text{L}(\widetilde{t})$, which is identical
to the negative of the reduced excess number of particles per cross-sectional
area in the liquid (see Eq.~\Eq{GammaL2}).
The asymptotic behavior is given by 
$-\widetilde{\Gamma}_\text{L}(\widetilde{t}\to\infty)\sim\widetilde{t}^{1/2}$.
Therefore, the number of evaporated particles is not bounded, but the rate
of evaporation is approaching zero.

Figure~\ref{fig:evaporation}(b) depicts the absolute reduced interface
position $|\widetilde{z}_{1/2}(\widetilde{t})|$, which is defined by the
location of the occupancy $50\%$, i.e., by
$\widetilde{\phi}(\widetilde{z}_{1/2}(\widetilde{t}),\widetilde{t})=1/2$.
One observes 
$|\widetilde{z}_{1/2}(\widetilde{t}\to\infty)|\sim\widetilde{t}^{1/2}$, which
implies that the interface is shifted infinitely far from the location of the
Gibbs dividing interface.
It is interesting to note that $\widetilde{z}_{1/2}(\widetilde{t})$ is
positive, i.e., the interface is shifted towards the vapor phase, for
$(\Delta\widetilde{\mu}_\text{L},\Delta\widetilde{\mu}_\text{V})=(1,-1)$ and
$(1,0)$, whereas it is negative, i.e., the interface is shifted towards the
liquid phase, for 
$(\Delta\widetilde{\mu}_\text{L},\Delta\widetilde{\mu}_\text{V})=(0,-1)$.
This finding may be illustrated by saying that an oversaturated liquid
is ``flooding'' the vapor phase, whereas an undersaturated vapor is ``eroding''
the liquid phase.


\section{\label{sec:conclusions}Conclusions and Summary}

In this work the early-stage relaxation of equilibrium and non-equilibrium 
interfaces in colloidal dispersions have been studied, which can separate in a 
colloidal-rich (liquid) phase and a colloidal-poor (vapor) phase 
(Fig.~\ref{fig:pd}).
Close to the critical point the interface formation is considerably slowed
down as compared to temperatures further away (Fig.~\ref{fig:profiles}).
Moreover, for the bulk phases being not at equilibrium the process of
interface formation is superimposed by the evaporation of colloidal particles
from the liquid into the vapor phase (Fig.~\ref{fig:profiles}).
However, the most surprising observation is that, irrespective of how much the 
bulk phases differ from two-phase coexistence, the interfacial structure 
approaches that at two-phase coexistence during the early-stage relaxation
process.
This surprising observation implies that the relaxation towards
global equilibrium of the interface is not following but preceding that of
the bulk phases, i.e., the interface relaxes independently of the bulk phases.

During the early-stage relaxation process the local chemical potential and the 
flux, which, as functions of position, interpolate between the corresponding 
bulk values, approach scaling forms (Fig.~\ref{fig:muj} and Eqs.~\Eq{muscaling}
and \Eq{jscaling}).
On the one hand, the leading order contributions in this scaling
exhibit power-law behavior in time, the values of the exponents of which depend
on whether an equilibrium or a non-equilibrium system is considered 
(Eqs.~\Eq{muscaling} and \Eq{jscaling}).
The occurrence of power-law behavior is linked to the underlying conserved
dynamics (model B), while a non-conserved dynamics (e.g., model A 
\cite{Hohenberg1977}) would give rise to an exponential decay.
On the other hand, the spatial range in which the local chemical potential
interpolates between the bulk values grows diffusively with time so that the
gradient, and thus the flux, decays with time.
Hence, the chemical potential becomes locally constant during the early-stage
relaxation process, which, at the value of the coexistence chemical potential, 
explains the occurrence of a liquid-vapor interfacial structure identical to 
that for two-phase coexistence, even in non-equilibrium systems.
Consequently, the interfacial tension decays to the value at two-phase
coexistence irrespective of whether an equilibrium or a non-equilibrium
system is considered (Fig.~\ref{fig:gammalin}), while the degree of 
non-equilibrium merely determines the quantitative deviation from the
equilibrium value.

Concerning the questions raised in the Introduction on non-equilibrium
interfaces in molecular fluids one notes that these systems, in contrast to
colloidal fluids, have to be described by model H dynamics 
\cite{Hohenberg1977} rather than by model B dynamics. 
Hence, due to the possibility of waves and turbulence within model H dynamics,
it is conceivable that the decay modes of the interfacial tensions
(Fig.~\ref{fig:gammalog}) and of the dissipation rates (Fig.~\ref{fig:P})
could be different for molecular and for colloidal fluids.
However, due to the line of arguments for model B dynamics given above, which
merely relies on the existence of a local chemical potential which interpolates
between the bulk values and which becomes locally constant, the 
interfacial tensions in non-equilibrium systems within model H dynamics can 
also be expected to approach the coexistence values during the early-stage
relaxation process.

Finally, the evaporation of colloidal particles from the liquid into the vapor
phase in non-equilibrium systems exhibits diffusive signatures 
(Fig.~\ref{fig:evaporation}), which is obviously linked to the underlying 
conserved dynamics.

To summarize the present work, the early-stage relaxation of an 
interface in non-equilibrium colloidal fluids, during which the interface but
not the bulk phases relax, towards that for two-phase coexistence has been
observed, explained and quantified within a simple model.
It has been argued that during the early-stage process an approach of 
quantities such as the interfacial tension towards the equilibrium values can
be expected to occur also for molecular fluids, whose dynamics differ from 
that of colloidal fluids.
Experimental investigations using colloidal fluids may be interesting to verify
the theoretical picture obtained here.


\begin{acknowledgments}
We like to thank S.\ Dietrich, R.\ Evans, M.\ Kr\"{u}ger, A.\ Macio{\l}ek and 
P.\ Teixeira for helpful comments.
\end{acknowledgments}



\end{document}